# Near Universal Consistency of the Maximum Pseudolikelihood Estimator for Discrete Models

Hien D. Nguyen*

August 30, 2017

**Abstract**

Maximum pseudolikelihood (MPL) estimators are useful alternatives to maximum likelihood (ML) estimators when likelihood functions are more difficult to manipulate than their marginal and conditional components. Furthermore, MPL estimators subsume a large number of estimation techniques including ML estimators, maximum composite marginal likelihood estimators, and maximum pairwise likelihood estimators. When considering only the estimation of discrete models (on a possibly countably infinite support), we show that a simple finiteness assumption on an entropy-based measure is sufficient for assessing the consistency of the MPL estimator. As a consequence, we demonstrate that the MPL estimator of any discrete model on a bounded support will be consistent. Our result is valid in parametric, semiparametric, and nonparametric settings.

## 1 Introduction

Let $\boldsymbol{X} \in \mathbb{X}$ be a random variable that takes values over the domain $\mathbb{X}$. Suppose that $\boldsymbol{X}$ arises from some probability model with density function $f_0(\boldsymbol{x})$. One of the core problem domains of machine learning, signal processing, and statistics is to devise an estimator $\hat{f}_n$ for $f_0$ that is in some sense close, using an independent and identically distributed (IID) sample $\mathbf{X}_n = \{\boldsymbol{X}_i\}_{i=1}^n$ from a distribution with density $f_0$. In such estimation problems, the general measure of success is consistency (i.e. for $\hat{f}_n$ to approach $f_0$ in some sense, as $n$ goes to infinity). The importance and fundamental nature of consistency is expounded well in Vapnik (2000, Ch. 2).

*Department of Mathematics and Statistics, La Trobe University, Melbourne Australia. (Email: h.nguyen5@latrobe.edu.au)

Let $f = f(\cdot;\boldsymbol{\theta})$ be parametrically determined family of probability density functions (PDFs), where $\boldsymbol{\theta}_0 \in \Theta \subset \mathbb{R}^p$ for some $p \in \mathbb{N}$. Assuming that $f_0$ can be written as $f(\cdot;\boldsymbol{\theta}_0)$, for some $\boldsymbol{\theta}_0 \in \Theta$, the conditions for the consistency of the maximum likelihood (ML) estimator $\hat{\boldsymbol{\theta}}_n$, which maximizes the log-likelihood function $\sum_{i=1}^{n} \log f(\boldsymbol{X}_i;\boldsymbol{\theta})$, are well-established in foundational works such as Wald (1949) and Kiefer & Wolfowitz (1956). When we estimate $f_0$ using PDFs from some general (potentially nonparametric) class $\mathcal{F}$, conditions for the consistency of the ML estimator

$$\hat{f}_n = \arg\max_{f\in\mathcal{F}} \sum_{i=1}^{n} \log f(\boldsymbol{X}_i)$$

have also been established in works such as van de Geer (1993) and Patilea (2001); see also Gine & Nickl (2015, Sec. 7.2). Apart from the likelihood function, general consistency conditions for estimates obtained via extremum estimation (Amemiya, 1985, Ch. 4), empirical-risk minimization (Vapnik, 2000, Ch. 2), and minimum-contrast estimation (Bickel & Doksum, 2000, Sec. 2.2) have also been broadly studied.

The maximum pseudolikelihood (MPL) estimator is an extremum estimator that is related to the ML estimator. Let $\mathbb{X} \subset \mathbb{R}^q$ for $q \in \mathbb{N}$, and write $\boldsymbol{X}^\top = (X_1, ..., X_q)$ and $\boldsymbol{X}_i^\top = (X_{i1}, ..., X_{iq})$, for each $i \in [n]$ ($[n] = \{1, ..., n\}$). Define $\mathbb{S} = 2^{[q]} \setminus \{\mathbf{0}\}$ to be the power set of $[q]$ that excludes the zero-string $\mathbf{0}$ and define $f^S(\boldsymbol{x}^S)$ to be the marginal PDF over the coordinates of $\boldsymbol{X}$ that are in $S \in \mathbb{S}$. We write $\boldsymbol{X}^S$ as the vector that contains only the coordinates that are in the set $S$. Next, define $\mathbb{T}$ to be the partitions of all subsets of $[q]$ into two non-empty sets. Referring to the two partitioning as left and right, we write $\boldsymbol{X}^{\overleftarrow{T}}$ and $\boldsymbol{X}^{\overrightarrow{T}}$ to indicate the vector containing the elements of $\boldsymbol{X}$ that are selected in the left and right set, respectively. Let $f^T\left(\boldsymbol{x}^{\overleftarrow{T}}|\boldsymbol{x}^{\overrightarrow{T}}\right)$ be the conditional PDF of $\boldsymbol{X}^{\overleftarrow{T}}$ given $\boldsymbol{X}^{\overrightarrow{T}} = \boldsymbol{x}^{\overrightarrow{T}}$, for each $T$. Note that we use the usual convention of upper case for random variables and lower case for realizations.

For each set of constants $\boldsymbol{c} = \{c_S\}_{S\in\mathbb{S}} \in [0,\infty)^{|\mathbb{S}|}$ and $\boldsymbol{d} = \{d_T\}_{T\in\mathbb{T}} \in [0,\infty)^{|\mathbb{T}|}$ for which $\boldsymbol{c} \neq \mathbf{0}$ if $\boldsymbol{d} = \mathbf{0}$ and $\boldsymbol{d} \neq \mathbf{0}$ if $\boldsymbol{c} = \mathbf{0}$, we can define a pseudolikelihood (PL) function as

$$L_f^{\boldsymbol{cd}}(\mathbf{X}_n) = \sum_{i=1}^{n}\sum_{S\in\mathbb{S}} c_S \log f^S\left(\boldsymbol{X}_i^S\right) + \sum_{i=1}^{n}\sum_{T\in\mathbb{T}} d_T \log f^T\left(\boldsymbol{X}_i^{\overleftarrow{T}}|\boldsymbol{X}_i^{\overrightarrow{T}}\right), \quad (1)$$

where $\boldsymbol{X}_i^S$ is distributed with marginal PDF $f^S\left(\boldsymbol{x}^S\right)$, and $\boldsymbol{X}_i^{\overleftarrow{T}}$ and $\boldsymbol{X}_i^{\overrightarrow{T}}$ are distributed with marginal PDF $f^T\left(\boldsymbol{x}^{\overleftarrow{T}}|\boldsymbol{x}^{\overrightarrow{T}}\right)f^{\overrightarrow{T}}\left(\boldsymbol{x}^{\overrightarrow{T}}\right)$, as per their counterparts without subscripts: $\boldsymbol{X}^S$, $\boldsymbol{X}^{\overleftarrow{T}}$, and $\boldsymbol{X}^{\overrightarrow{T}}$; see Arnold & Strauss (1991) for a definition of the PL function that is compatible with Equation (1). The PL functions of form (1) can also be viewed as the nonparametric generalization of the composite likelihood functions that were studied in Lindsay (1988) and Cox & Reid (2004). We note that $|\mathbb{S}| = 2^q - 1$ and $|\mathbb{T}|$ can be counted using Stirling numbers of the second kind; see Charalambides (2002, Ch. 8) for details.

We say that

$$\tilde{f}_n \in \left\{\tilde{f} \in \mathcal{F}:\ L_{\tilde{f}}^{\boldsymbol{cd}}\left(\mathbf{X}_n\right) = \max_{f\in\mathcal{F}} L_f^{\boldsymbol{cd}}\left(\mathbf{X}_n\right)\right\},$$

is the MPL estimator over the functional class $\mathcal{F}$, provided that all of the necessary marginals $\tilde{f}_n^S$ and conditionals $\tilde{f}_n^T\left(\cdot|\cdot\right)$ that contribute to $L_f^{\boldsymbol{cd}}$ (i.e. $c_S \neq 0$ or $d_T \neq 0$) are compatible with $\tilde{f}_n$, in the sense that we can compute all $\tilde{f}_n^S$ and $\tilde{f}_n^T\left(\cdot|\cdot\right)$ from $\tilde{f}_n$ using the usual laws for deriving marginal and conditional PDFs; see Joe (1997) and Arnold et al. (1999) for further details regarding the theory of compatibility when defining PL functions. When $\mathcal{F}$ is a parametric family such that $f_0 = f\left(x;\boldsymbol{\theta}_0\right)$, and when all of the marginal and conditional densities $f^S = f^S\left(\cdot;\boldsymbol{\theta}\right)$ and $f^T\left(\cdot|\cdot\right) = f^T\left(\cdot|\cdot;\boldsymbol{\theta}\right)$ (for $c_S \neq 0$ and $d_T \neq 0$) are all compatible with some density $f\left(\cdot;\boldsymbol{\theta}\right)$, for $\boldsymbol{\theta} \in \Theta$, the conditions for consistency for the parametric MPL estimator

$$\tilde{\boldsymbol{\theta}}_n \in \left\{\tilde{\boldsymbol{\theta}} \in \Theta:\ L_f^{\boldsymbol{cd}}\left(\mathbf{X}_n;\tilde{\boldsymbol{\theta}}\right) = \max_{\boldsymbol{\theta}\in\Theta} L_f^{\boldsymbol{cd}}\left(\mathbf{X}_n;\boldsymbol{\theta}\right)\right\}$$

have been obtained by Lindsay (1988) and Arnold & Strauss (1991); see also Molenberghs & Verbeke (2005, Ch. 9). Here, $L_f^{\boldsymbol{cd}}\left(\mathbf{X}_n;\boldsymbol{\theta}\right)$ is defined as per (1), with $f^S\left(\cdot;\boldsymbol{\theta}\right)$ and $f^T\left(\cdot|\cdot;\boldsymbol{\theta}\right)$ replacing $f^S$ and $f^T\left(\cdot|\cdot\right)$, respectively, for each $S$ and $T$.

MPL estimation, as an alternative to ML estimation, was first introduced in Besag (1974). The use of MPL estimation has become ubiquitous in statistics and machine learning, with applications from multivariate modeling of toxicology data (Geys et al., 1999), to genetic mapping (Ahfock et al., 2014), and fitting of neural networks (Hyvarinen, 2006; Nguyen & Wood, 2016a,b). Numerous other applications are presented in Arnold et al. (1999) and Molenberghs & Verbeke (2005). Recent reviews of MPL estimation appear in Varin et al. (2011) and Yi (2017).

We now concentrate our attention to the estimation of probability mass functions (PMFs) for random variables in $\mathbb{X} \subset \mathbb{Z}^q$, where $\mathbb{X}$ can potentially be countable infinite. When $q = 1$, Seo & Lindsay (2013) presented a simple criterion for checking the consistency ML estimator for any PMF over $\mathbb{Z}$ via a simple entropy criterion. Using their criterion, it is then easy to observe that if $\mathbb{X}$ is bounded, then the criterion is always satisfied and thus provides a universal consistency result. When $\mathbb{X}$ is unbounded, however, there still may exists classes that cannot be consistently estimated via MPL estimation.

Using the proof technique from Seo & Lindsay (2013), we derive an MPL estimator consistency theorem for the estimation of PMFs for data in $\mathbb{X} \subset \mathbb{Z}^q$. Our result subsumes that of Seo & Lindsay (2013) as the ML estimator, when $q = 1$, is an MPL estimator. Our result is suitable for the parametric, semiparametric, and nonparametric estimation settings. However, in the case of parametric and semiparametric estimation, it can only identify the parameters of interest to up to a set of maxima, unless the PL function is identifiable with respect to the parameter space. We proceed with the presentation of the main result in Section 2. Some example applications of the main result are provided in Section 3.

# 2 Main Result

Assume from here on that $\mathbb{X} \subset \mathbb{Z}^q$. Let $\mathbb{X}^S$, $\mathbb{X}^{\overleftarrow{T}}$, and $\mathbb{X}^{\overrightarrow{T}}$ be the coordinates of $\mathbb{X}$ that correspond to the sets of coordinates in $S \in \mathbb{S}$, $\overleftarrow{T}$, and $\overrightarrow{T}$, respectively, where $T \in \mathbb{T}$. Let the IID sample $\mathbf{X}_n$ arise from a distribution with PMF $f_0 \in \mathcal{F}$, where $\mathcal{F}$ is some class of PMFs, with marginals and conditions $f_0^S$ and $f_0^T(\cdot|\cdot)$ for all $S$ and $T$. If all relevant marginal and conditional PMFs $\tilde{f}_n^S$ and $\tilde{f}_n^T(\cdot|\cdot)$ (i.e. $c_S \neq 0$ and $d_T \neq 0$) are compatible with the estimator $\tilde{f}_n$, and if $\tilde{f}_n^S(\boldsymbol{x}^S) \to f_0^S(\boldsymbol{x}^S)$ and $\tilde{f}_n^T\left(\boldsymbol{x}^{\overleftarrow{T}}|\boldsymbol{x}^{\overrightarrow{T}}\right) \to f_0^T\left(\boldsymbol{x}^{\overleftarrow{T}}|\boldsymbol{x}^{\overrightarrow{T}}\right)$ with probability 1, for every $\boldsymbol{x} \in \mathbb{X}$, $\boldsymbol{x}^S \in \mathbb{X}^S$, $\boldsymbol{x}^{\overleftarrow{T}} \in \mathbb{X}^{\overleftarrow{T}}$, and $\boldsymbol{x}^{\overrightarrow{T}} \in \mathbb{X}^{\overrightarrow{T}}$, then we say that $\tilde{f}_n$ is a consistent estimator for $f_0$.

*Remark* 1. Note that we do not include the condition that $\tilde{f}_n(\boldsymbol{x}) \to f_0(\boldsymbol{x})$, with probability 1, as a condition of our definition of consistency (unless $c_{S'} = 1$, where $S'$ is the set of all coordinates of $\mathbb{X}$). This is because it may be possible that there are more than one PMFs $\tilde{f}_n$ that may be compatible with the marginal and conditional PMFs in the sets $\mathbb{S}$ and $\mathbb{T}$. This is the problem of nonidentifiability (cf. Yi, 2017). Thus, we can consider this mode of consistency as the existence of an estimator $\tilde{f}_n \in \mathcal{F}$ having compatible relevant marginal and conditional PMFs that converge with probability 1 to the marginals of the generative process $f_0$.

Define the pseudo-entropy of any $f \in \mathcal{F}$ as

$$\begin{aligned} H(f) &= \sum_{S \in \mathbb{S}} \sum_{\boldsymbol{x}^S \in \mathbb{X}^S} -c_S f^S\left(\boldsymbol{x}^S\right) \log f^S\left(\boldsymbol{x}^S\right) \\ &\quad + \sum_{T \in \mathbb{T}} d_T \sum_{\boldsymbol{x}^{\overrightarrow{T}} \in \mathbb{X}^{\overrightarrow{T}}} \sum_{\boldsymbol{x}^{\overleftarrow{T}} \in \mathbb{X}^{\overleftarrow{T}}} -f^T\left(\boldsymbol{x}^{\overleftarrow{T}} | \boldsymbol{x}^{\overrightarrow{T}}\right) \log f^T\left(\boldsymbol{x}^{\overleftarrow{T}} | \boldsymbol{x}^{\overrightarrow{T}}\right). \end{aligned}$$

Let $\mu$ be a sub-probability measure (SPM) on $\mathbb{R}^r$ (in the sense that $\mu(\mathbb{R}^r) \leq 1$; $r \in \mathbb{N}$) and let $\mathcal{G}$ be the class of continuous functions that are supported on some compact subset of $\mathbb{R}^r$. We say that a sequence $\{\mu_n\}$ of SPMs converges vaguely to $\mu$ if $\mu_n g \to \mu g$, as $n \to \infty$, for every $g \in \mathcal{G}$ (cf. Kallenberg, 1997, Ch. 7). We denote vague convergence as $\mu_n \overset{v}{\to} \mu$ and shall use the following result from Chung (2001) in order to obtain our consistency theorem; see also Bauer (2001, Sec. 31).

**Lemma 1** (Chung, 2001, Thm. 4.3.4)**.** *If every vaguely convergent subsequence of* $\{\mu_n\}$ *convergences to the same limit* $\mu$*, then* $\mu_n \overset{v}{\to} \mu$.

**Theorem 1.** *Let* $\mathbf{X}_n$ *be an IID sample that arises from a distribution with PMF* $f_0 \in \mathcal{F}$*. If* $H(f_0)$ *is finite and if all relevant marginal and conditional PMFs* $\tilde{f}_n^S$ *and* $\tilde{f}_n^T$ *are compatible with* $\tilde{f}_n \in \mathcal{F}$ *(i.e.* $c_S \neq 0$ *or* $d_T \neq 0$*, for* $S \in \mathbb{S}$ *and* $T \in \mathbb{T}$*), then the MPL estimator* $\tilde{f}_n \in \mathcal{F}$ *is a consistent estimator for* $f_0$*.*

*Proof.* Fix a realization $\mathbf{x}_n = \{\boldsymbol{x}_n\}$ of $\mathbf{X}_n$ and let $n_1, n_2, ..., n_m$be a subsequence of $[n]$. Let $\tilde{f}_m$ be the MPL estimator obtained from the subsequence and assume that each necessary $\tilde{f}_m^S$ and $\tilde{f}_m^T\left(\cdot|\boldsymbol{x}^{\overrightarrow{T}}\right)$ converges vaguely to $\tilde{f}^S$ and $\tilde{f}^T\left(\cdot|\boldsymbol{x}^{\overrightarrow{T}}\right)$, respectively. We shall write $\sum_k$ for $\sum_{k \in \{n_1, \dots, n_m\}}$. First rewrite the PL function (1), over the subsequence, as

$$\sum_k \sum_{S \in \mathbb{S}} c_S \sum_{\boldsymbol{x}^S \in \mathbb{X}^S} I\left(\boldsymbol{x}_k^S = \boldsymbol{x}^S\right) \log f^S\left(\boldsymbol{x}_k^S\right) + \sum_k \sum_{T \in \mathbb{T}} d_T \sum_{\boldsymbol{x}^{\overrightarrow{T}} \in \mathbb{X}^{\overrightarrow{T}}} \sum_{\boldsymbol{x}^{\overleftarrow{T}} \in \mathbb{X}^{\overleftarrow{T}}} I\left(\boldsymbol{x}_k^{\overleftarrow{T}} = \boldsymbol{x}^{\overleftarrow{T}}, \boldsymbol{x}_k^{\overrightarrow{T}} = \boldsymbol{x}^{\overrightarrow{T}}\right) \log f^T\left(\boldsymbol{x}^{\overleftarrow{T}} | \boldsymbol{x}^{\overrightarrow{T}}\right),$$

where $I(A)$ be an indicator function that takes value 1 if proposition $A$ is true and 0 otherwise. By definition of the MPL estimator, we must have

$$
\begin{aligned}
0 &\geq \liminf_{m\to\infty} \frac{1}{m} \sum_k \sum_{S\in\mathbb{S}} c_S I\left(\boldsymbol{x}_k^S = \boldsymbol{x}^S\right) \log\left[\frac{f_0^S\left(\boldsymbol{x}^S\right)}{\tilde{f}_m^S\left(\boldsymbol{x}^S\right)}\right] \\
&+\liminf_{m\to\infty} \frac{1}{m} \sum_k \sum_{T\in\mathbb{T}} d_T \sum_{\boldsymbol{x}^{\overrightarrow{T}}\in\mathbb{X}^{\overrightarrow{T}}} \sum_{\boldsymbol{x}^{\overleftarrow{T}}\in\mathbb{X}^{\overleftarrow{T}}} I\left(\boldsymbol{x}_k^{\overleftarrow{T}} = \boldsymbol{x}^{\overleftarrow{T}}, \boldsymbol{x}_k^{\overrightarrow{T}} = \boldsymbol{x}^{\overrightarrow{T}}\right) \log\left[\frac{f_0^T\left(\boldsymbol{x}^{\overleftarrow{T}}|\boldsymbol{x}^{\overrightarrow{T}}\right)}{\tilde{f}_m^T\left(\boldsymbol{x}^{\overleftarrow{T}}|\boldsymbol{x}^{\overrightarrow{T}}\right)}\right] \\
&= \liminf_{m\to\infty} \sum_{S\in\mathbb{S}} c_S \sum_{\boldsymbol{x}^S\in\mathbb{X}^S} \sum_k \frac{I\left(\boldsymbol{x}_k^S = \boldsymbol{x}^S\right)}{m} \log\left[\frac{f_0^S\left(\boldsymbol{x}^S\right)}{\tilde{f}_m^S\left(\boldsymbol{x}^S\right)}\right] \\
&+\liminf_{m\to\infty} \sum_{T\in\mathbb{T}} d_T \sum_{\boldsymbol{x}^{\overrightarrow{T}}\in\mathbb{X}^{\overrightarrow{T}}} \sum_{\boldsymbol{x}^{\overleftarrow{T}}\in\mathbb{X}^{\overleftarrow{T}}} \sum_k \frac{I\left(\boldsymbol{x}_k^{\overleftarrow{T}} = \boldsymbol{x}^{\overleftarrow{T}}, \boldsymbol{x}_k^{\overrightarrow{T}} = \boldsymbol{x}^{\overrightarrow{T}}\right)}{m} \log\left[\frac{f_0^T\left(\boldsymbol{x}^{\overleftarrow{T}}|\boldsymbol{x}^{\overrightarrow{T}}\right)}{\tilde{f}_m^T\left(\boldsymbol{x}^{\overleftarrow{T}}|\boldsymbol{x}^{\overrightarrow{T}}\right)}\right] \\
&= \liminf_{m\to\infty} \sum_{S\in\mathbb{S}} c_S \sum_{\boldsymbol{x}^S\in\mathbb{X}^S} p_m^S\left(\boldsymbol{x}^S\right) \log\left[\frac{f_0^S\left(\boldsymbol{x}^S\right)}{\tilde{f}_m^S\left(\boldsymbol{x}^S\right)}\right] \\
&+\liminf_{m\to\infty} \sum_{T\in\mathbb{T}} d_T \sum_{\boldsymbol{x}^{\overrightarrow{T}}\in\mathbb{X}^{\overrightarrow{T}}} \sum_{\boldsymbol{x}^{\overleftarrow{T}}\in\mathbb{X}^{\overleftarrow{T}}} p_m^T\left(\boldsymbol{x}^{\overleftarrow{T}}, \boldsymbol{x}^{\overrightarrow{T}}\right) \log\left[\frac{f_0^T\left(\boldsymbol{x}^{\overleftarrow{T}}|\boldsymbol{x}^{\overrightarrow{T}}\right)}{\tilde{f}_m^T\left(\boldsymbol{x}^{\overleftarrow{T}}|\boldsymbol{x}^{\overrightarrow{T}}\right)}\right] \\
&\geq \liminf_{m\to\infty} \sum_{S\in\mathbb{S}} d_S \sum_{\boldsymbol{x}^S\in\mathbb{X}^S} p_m^S \log f_0^S\left(\boldsymbol{x}^S\right) \\
&+\liminf_{m\to\infty} \sum_{T\in\mathbb{T}} d_T \sum_{\boldsymbol{x}^{\overrightarrow{T}}\in\mathbb{X}^{\overrightarrow{T}}} \sum_{\boldsymbol{x}^{\overleftarrow{T}}\in\mathbb{X}^{\overleftarrow{T}}} p_m^T\left(\boldsymbol{x}^{\overleftarrow{T}}, \boldsymbol{x}^{\overrightarrow{T}}\right) \log f_0^T\left(\boldsymbol{x}^{\overleftarrow{T}}|\boldsymbol{x}^{\overrightarrow{T}}\right) \\
&+\liminf_{m\to\infty} \sum_{S\in\mathbb{S}} d_S \sum_{\boldsymbol{x}^S\in\mathbb{X}^S} -p_m^S\left(\boldsymbol{x}^S\right) \log \tilde{f}_m^S\left(\boldsymbol{x}^S\right) \\
&+\liminf_{m\to\infty} \sum_{T\in\mathbb{T}} d_T \sum_{\boldsymbol{x}^{\overrightarrow{T}}\in\mathbb{X}^{\overrightarrow{T}}} \sum_{\boldsymbol{x}^{\overleftarrow{T}}\in\mathbb{X}^{\overleftarrow{T}}} -p_m^T\left(\boldsymbol{x}^{\overleftarrow{T}}, \boldsymbol{x}^{\overrightarrow{T}}\right) \log \tilde{f}_m^T\left(\boldsymbol{x}^{\overleftarrow{T}}|\boldsymbol{x}^{\overrightarrow{T}}\right),
\end{aligned}
$$

where $p_m^S\left(\boldsymbol{x}^S\right) = m^{-1}\sum_k I\left(\boldsymbol{s}_k^S = \boldsymbol{s}_k\right)$ and $p_m^T\left(\boldsymbol{x}^{\overleftarrow{T}}, \boldsymbol{x}^{\overrightarrow{T}}\right) = m^{-1}\sum_k I\left(\boldsymbol{x}_k^{\overleftarrow{T}} = \boldsymbol{x}^{\overleftarrow{T}}, \boldsymbol{x}_k^{\overrightarrow{T}} = \boldsymbol{x}^{\overrightarrow{T}}\right)$, are sample proportions for observing the relevant events.

Since all $c_S$ and $d_T$ are non-negative, and all $-\log \tilde{f}_m^S\left(\boldsymbol{x}^S\right)$ and $-\log \tilde{f}_m^T\left(\boldsymbol{x}^{\overleftarrow{T}}|\boldsymbol{x}^{\overrightarrow{T}}\right)$ are non-negative, we can apply Fatou's Lemma to swap $\liminf$ and the sum in the last two terms and maintain the chain of inequalities.

That is, the last expression above is greater than or equal to

$$\begin{aligned}
&\liminf_{m\to\infty}\sum_{S\in\mathbb{S}}c_S\sum_{\boldsymbol{x}^S\in\mathbb{X}^S}p_m^S\left(\boldsymbol{x}^S\right)\log f_0^S\left(\boldsymbol{x}^S\right)\\
&+\liminf_{m\to\infty}\sum_{T\in\mathbb{T}}d_T\sum_{\boldsymbol{x}^{\overrightarrow{T}}\in\mathbb{X}^{\overrightarrow{T}}}\sum_{\boldsymbol{x}^{\overleftarrow{T}}\in\mathbb{X}^{\overleftarrow{T}}}p_m^T\left(\boldsymbol{x}^{\overleftarrow{T}},\boldsymbol{x}^{\overrightarrow{T}}\right)\log f_0^T\left(\boldsymbol{x}^{\overleftarrow{T}}|\boldsymbol{x}^{\overrightarrow{T}}\right)\\
&+\sum_{S\in\mathbb{S}}c_S\sum_{\boldsymbol{x}^S\in\mathbb{X}^S}\liminf_{m\to\infty}-p_m^S\left(\boldsymbol{x}^S\right)\log\tilde{f}_m^S\left(\boldsymbol{x}^S\right)\\
&+\sum_{T\in\mathbb{T}}d_T\sum_{\boldsymbol{x}^{\overrightarrow{T}}\in\mathbb{X}^{\overrightarrow{T}}}\sum_{\boldsymbol{x}^{\overleftarrow{T}}\in\mathbb{X}^{\overleftarrow{T}}}\liminf_{m\to\infty}-p_m^T\left(\boldsymbol{x}^{\overleftarrow{T}},\boldsymbol{x}^{\overrightarrow{T}}\right)\log\tilde{f}_m^T\left(\boldsymbol{x}^{\overleftarrow{T}}|\boldsymbol{x}^{\overrightarrow{T}}\right).
\end{aligned}$$

Finiteness of $H(f_0)$ implies the existence of the first two limits. Further, all proportions converge under the strong law of large numbers to their corresponding limits under the generative model. That is $p_m^S\left(\boldsymbol{x}^S\right)\to f_0^S\left(\boldsymbol{x}^S\right)$ and $p_m^T\left(\boldsymbol{x}^{\overleftarrow{T}},\boldsymbol{x}^{\overrightarrow{T}}\right)\to f_0^T\left(\boldsymbol{x}^{\overleftarrow{T}}|\boldsymbol{x}^{\overrightarrow{T}}\right)f_0^{\overrightarrow{T}}\left(\boldsymbol{x}^{\overrightarrow{T}}\right)$, almost surely as $n\to\infty$. Next, under our hypothesis, the subsequences $\tilde{f}_m^S$ and $\tilde{f}_m^T(\cdot|\cdot)$ converge to SPMs $f^S$ and $f^T(\cdot|\cdot)$, respectively. Upon substitution of all of our observations, we obtain the equality, with probability 1, of the last line above to the expression

$$\begin{aligned}
&\sum_{S\in\mathbb{S}}c_S\sum_{\boldsymbol{x}^S\in\mathbb{X}^S}f_0^S\left(\boldsymbol{x}^S\right)\log f_0^S\left(\boldsymbol{x}^S\right)\\
&+\sum_{T\in\mathbb{T}}d_T\sum_{\boldsymbol{x}^{\overrightarrow{T}}\in\mathbb{X}^{\overrightarrow{T}}}f_0^{\overrightarrow{T}}\left(\boldsymbol{x}^{\overrightarrow{T}}\right)\sum_{\boldsymbol{x}^{\overleftarrow{T}}\in\mathbb{X}^{\overleftarrow{T}}}f_0^T\left(\boldsymbol{x}^{\overleftarrow{T}}|\boldsymbol{x}^{\overrightarrow{T}}\right)\log f_0^T\left(\boldsymbol{x}^{\overleftarrow{T}}|\boldsymbol{x}^{\overrightarrow{T}}\right)\\
&-\sum_{S\in\mathbb{S}}c_S\sum_{\boldsymbol{x}^S\in\mathbb{X}^S}f_0^S\left(\boldsymbol{x}^S\right)\log f^S\left(\boldsymbol{x}^S\right)\\
&-\sum_{T\in\mathbb{T}}d_T\sum_{\boldsymbol{x}^{\overrightarrow{T}}\in\mathbb{X}^{\overrightarrow{T}}}f_0^{\overrightarrow{T}}\left(\boldsymbol{x}^{\overrightarrow{T}}\right)\sum_{\boldsymbol{x}^{\overleftarrow{T}}\in\mathbb{X}^{\overleftarrow{T}}}f_0^T\left(\boldsymbol{x}^{\overleftarrow{T}}|\boldsymbol{x}^{\overrightarrow{T}}\right)\log f^T\left(\boldsymbol{x}^{\overleftarrow{T}}|\boldsymbol{x}^{\overrightarrow{T}}\right)\\
=\;&\sum_{S\in\mathbb{S}}c_S\sum_{\boldsymbol{x}^S\in\mathbb{X}^S}f_0^S\left(\boldsymbol{x}^S\right)\log\left[\frac{f_0^S\left(\boldsymbol{x}^S\right)}{f^S\left(\boldsymbol{x}^S\right)}\right]\\
&+\sum_{T\in\mathbb{T}}d_T\sum_{\boldsymbol{x}^{\overrightarrow{T}}\in\mathbb{X}^{\overrightarrow{T}}}f_0^{\overrightarrow{T}}\left(\boldsymbol{x}^{\overrightarrow{T}}\right)\sum_{\boldsymbol{x}^{\overleftarrow{T}}\in\mathbb{X}^{\overleftarrow{T}}}f_0^T\left(\boldsymbol{x}^{\overleftarrow{T}}|\boldsymbol{x}^{\overrightarrow{T}}\right)\log\left[\frac{f_0^T\left(\boldsymbol{x}^{\overleftarrow{T}}|\boldsymbol{x}^{\overrightarrow{T}}\right)}{f^T\left(\boldsymbol{x}^{\overleftarrow{T}}|\boldsymbol{x}^{\overrightarrow{T}}\right)}\right]\geq 0,
\end{aligned}$$

where the final inequality is due to the log-sum inequality along with the fact that each $f^S$ and $f^T(\cdot|\cdot)$ is a SPM, and the fact that all $c_S$, $d_T$, and $f_0^{\overrightarrow{T}}$ are non-negative. Thus, we observe that every line in the the string of inequalities equates to zero, which can only be the case if $f^S=f_0^S$ and $f^T\left(\cdot|\boldsymbol{x}^{\overrightarrow{T}}\right)=f_0^T\left(\cdot|\boldsymbol{x}^{\overrightarrow{T}}\right)$, for all $S$ and $T$, and for all

elements of each domain (again, due to the log-sum inequality). Thus, we obtain the result that every subsequence of $\tilde{f}_n^S$, and $\tilde{f}_n^T\left(\cdot|\boldsymbol{x}^{\overrightarrow{T}}\right)$ (i.e. $\tilde{f}_m^S$ and $\tilde{f}_m^T\left(\cdot|\boldsymbol{x}^{\overrightarrow{T}}\right)$) converges vaguely to the same limit, $f^S$ and $f^T\left(\cdot|\boldsymbol{x}^{\overrightarrow{T}}\right)$. Thus, by Lemma 1, we have $\tilde{f}_n^S \overset{v}{\to} f^S = f_0^S$ and $\tilde{f}_n^T\left(\cdot|\boldsymbol{x}^{\overrightarrow{T}}\right) \overset{v}{\to} f^T\left(\cdot|\boldsymbol{x}^{\overrightarrow{T}}\right) = f_0^T\left(\cdot|\boldsymbol{x}^{\overrightarrow{T}}\right)$, for each $S$ and $T$. Since vague convergence implies the convergence of all continuous and compactly supported functions (e.g. a PMF on any point of support), almost surely, we have almost sure pointwise convergence of all marginal and conditional PMFs. Finally, we set $\tilde{f}_n \in \mathcal{F}$ to be a sequence of PMFs for which the marginals and conditions are compatible to complete the proof. □

*Remark* 2. Although the result is established in a nonparametric setting, it is not difficult to rephrase in terms of parametric models. For example, we can consider the class of functions $\mathcal{F}_{\boldsymbol{\theta}} = \{f(\cdot;\boldsymbol{\theta}) : \boldsymbol{\theta} \in \Theta\}$. In such as case, we can define the MPL estimator as $\tilde{f}_n = f\left(\cdot;\tilde{\boldsymbol{\theta}}_n\right)$, where $\tilde{\boldsymbol{\theta}}_n \in \Theta_n$ and

$$\Theta_n = \left\{\tilde{\boldsymbol{\theta}} \in \Theta : L_f^{\boldsymbol{cd}}\left(\mathbf{X}_n;\tilde{\boldsymbol{\theta}}\right) = \max_{\boldsymbol{\theta}\in\Theta} L_f^{\boldsymbol{cd}}\left(\mathbf{X}_n;\boldsymbol{\theta}\right)\right\}.$$

Here we assume that $L_f^{\boldsymbol{cd}}$ identifies all of the elements of $\boldsymbol{\theta}$ (cf. Yi, 2017). Suppose that the data $\mathbf{X}_n$ is generated from some $f_0 = f(x;\boldsymbol{\theta}_0)$ with marginal and conditional PMFs $f^S = f^S(\cdot;\boldsymbol{\theta}_0)$ and $f^T(\cdot|\cdot) = f^T(\cdot|\cdot;\boldsymbol{\theta}_0)$. Theorem 1 then establishes the fact that $f_n^S\left(\boldsymbol{x}^S;\tilde{\boldsymbol{\theta}}_n\right) \to f^S\left(\boldsymbol{x}^S;\boldsymbol{\theta}_0\right)$ and $f_n^T\left(\boldsymbol{x}^{\overleftarrow{T}}|\boldsymbol{x}^{\overrightarrow{T}};\tilde{\boldsymbol{\theta}}_n\right) \to f^T\left(\boldsymbol{x}^{\overleftarrow{T}}|\boldsymbol{x}^{\overrightarrow{T}};\boldsymbol{\theta}_0\right)$ with probability 1, for every $\boldsymbol{x} \in \mathbb{X}$, $\boldsymbol{x}^S \in \mathbb{X}^S$, $\boldsymbol{x}^{\overleftarrow{T}} \in \mathbb{X}^{\overleftarrow{T}}$, and $\boldsymbol{x}^{\overrightarrow{T}} \in \mathbb{X}^{\overrightarrow{T}}$. This convergence, however, does not guarantee that $\tilde{\boldsymbol{\theta}}_n$ will converge to $\boldsymbol{\theta}_0$ but only that $\inf_{\boldsymbol{\theta}\in\Theta_n}\|\boldsymbol{\theta}-\boldsymbol{\theta}_0\| \to 0$, almost surely; see the proof of Amemiya (1985, Thm. 4.1.2), for example. Further continuity assumptions on the function $f$ along with the identifiability of the mapping between $\boldsymbol{\theta}$ and $f(\cdot;\boldsymbol{\theta})$ is required in order to obtain the convergence of $\tilde{\boldsymbol{\theta}}_n$ to $\boldsymbol{\theta}_0$. For example, one can apply Kosorok (2008, Thm. 2.1.2).

*Remark* 3. Aside from the nonparametric (main) result, and the parametric result that is discussed in Remark 2, we can also obtain results regarding the MPL estimator for semiparametric models. Between the parametric and nonparametric cases, there are numerous ways to specify semiparametric models. A number of examples of semiparametric constructions are (i) mixing distributions, (ii) latent variable models, and (iii) basis-smoothed models. Examples of models in class (i)–(iii) are the finite mixture models that are considered in Johnson et al. (1997, Ch. 42, Sec. 2), the restricted Boltzmann machine of Smolensky (1986), and the polynomially-smoothed contingency tables of Jacob & Oliveira (2012), respectively.

In each of the semiparametric estimation scenarios above, the MPL estimator can be defined as the pair

$$\left(\tilde{f}, \tilde{\boldsymbol{\theta}}_n^\top\right) \in \left\{\tilde{f} \in \mathcal{F}, \tilde{\boldsymbol{\theta}} \in \Theta_{\tilde{f}} : \; L_{\tilde{f}}^{\boldsymbol{cd}}\left(\mathbf{X}_n; \tilde{\boldsymbol{\theta}}\right) = \max_{f \in \mathcal{F}} \max_{\boldsymbol{\theta} \in \Theta_f} L_f^{\boldsymbol{cd}}\left(\mathbf{X}_n\right)\right\}, \tag{2}$$

where each joint PMF $f(\cdot; \boldsymbol{\theta}) \in \mathcal{F}$ is a parametric model with parameter vector in the space $\Theta_f$, depending on $f(\cdot; \boldsymbol{\theta})$. Here, the expression $L_{\tilde{f}}^{\boldsymbol{cd}}\left(\boldsymbol{X}_n; \tilde{\boldsymbol{\theta}}\right)$ is as per its appearance in Remark 2, and each $f \in \mathcal{F}$ has the necessary marginal and conditional PMFs in order to define $L_{\tilde{f}}^{\boldsymbol{cd}}\left(\boldsymbol{X}_n; \tilde{\boldsymbol{\theta}}\right)$. Thus, MPL estimator (2) retrieves the model $f(\cdot; \boldsymbol{\theta}) \in \mathcal{F}$ that has the maximal log-PL value, subject to the parameter space $\Theta_f$. As in Remark 2, if the data $\mathbf{X}_n$ is generated from some $f(\cdot; \boldsymbol{\theta}_0) \in \mathcal{F}$ with marginal and conditional PMFs $f^S(\cdot; \boldsymbol{\theta}_0)$ and $f^T(\cdot|\cdot; \boldsymbol{\theta}_0)$, then Theorem 1 establishes the fact that $\tilde{f}_n^S\left(\boldsymbol{x}^S; \tilde{\boldsymbol{\theta}}_n\right) \to f^S\left(\boldsymbol{x}^S; \boldsymbol{\theta}_0\right)$ and $\tilde{f}_n^T\left(\boldsymbol{x}^{\overleftarrow{T}}|\boldsymbol{x}^{\overrightarrow{T}}; \tilde{\boldsymbol{\theta}}_n\right) \to f^T\left(\boldsymbol{x}^{\overleftarrow{T}}|\boldsymbol{x}^{\overrightarrow{T}}; \boldsymbol{\theta}_0\right)$ with probability 1, for every $\boldsymbol{x} \in \mathbb{X}$, $\boldsymbol{x}^S \in \mathbb{X}^S$, $\boldsymbol{x}^{\overleftarrow{T}} \in \mathbb{X}^{\overleftarrow{T}}$, and $\boldsymbol{x}^{\overrightarrow{T}} \in \mathbb{X}^{\overrightarrow{T}}$. As with the parametric and nonparametric cases, we must note that further assumptions are also required to establish the identifiability of the MPL estimator for semiparametric models.

*Remark* 4. For any $y \in [0, 1]$, we have

$$\max_y \; -y \log y = 1/e. \tag{3}$$

If $\mathbb{X}$ is a bounded set, then (3) implies that $H(f_0)$ will be finite as it is a finite sum of finite values, and thus the conditions for Theorem 1 will be met. This is the reason why we consider our result to be near universal.

*Remark* 5. As noted in the introduction, the MPL estimator subsumes the ML estimator (i.e., we can set $c_{S'} = 1$, where $S'$ is the set of all coordinates of $\mathbb{X}$, set $c_S = 0$ for all other $S \in \mathbb{S}$, and set $\boldsymbol{d} = \mathbf{0}$). Furthermore, the MPL estimator subsumes the composite marginal likelihood estimator of Varin (2008) as well as the pairwise likelihood estimator of Cox & Reid (2004) (i.e., only set $c_{S'} > 0$ for those $S'$ that contain a single coordinate of $\mathbb{X}$ or exactly two coordinates of $\mathbb{X}$, respectively, and set $\boldsymbol{d} = \mathbf{0}$). As such, the near universal consistency result of Theorem 1 also applies to these estimators.

## 3 Example Applications

**Example 1.** *(Categorical distribution)* We begin with a toy example. Let $\boldsymbol{X}_i$ be a random variable ($i \in [n]$) and suppose that $\mathbb{P}(\boldsymbol{X}_i = \boldsymbol{e}_k) = \pi_k > 0$, where $\boldsymbol{e}_k$ is a $q$-dimensional vector with a one at the $k$th coordinate and zeros at all other coordinates ($k \in [q]$), and where $\sum_{k=1}^{q} \pi_k = 1$. We say that $\boldsymbol{X}_i$ arises from a categorical distribution with parameter vector $\boldsymbol{\pi}^\top = (\pi_1, \dots, \pi_q)$, if its distribution can be characterized by the PMF

$$f(\boldsymbol{x}_i; \boldsymbol{\pi}) = \prod_{k=1}^{q} \pi_k^{I(\boldsymbol{x}_i = \boldsymbol{e}_k)}. \tag{4}$$

The categorical distribution is the single-trial case of the multinomial distribution and has numerous applications in machine learning and pattern recognition; see for example Bishop (2006, Ch. 9). The categorical distribution PMF (4) can be viewed as the least-restrictive nonparametric model over the space $\mathbb{X} = \{\boldsymbol{e}_k\}_{k=1}^{q}$.

Let $\boldsymbol{X}_i^\top = (x_{i1}, \dots, x_{iq})$ and let $\boldsymbol{X}_{(k)i}^\top = (X_1, \dots, X_{k-1}, X_{k+1}, \dots, X_q)$, for each $i \in [n]$ and $k \in [q]$. From (4), we can use the derivations of Albert & Denis (2012) to write the marginal PMFs

$$f\left(\boldsymbol{x}_{(q-1)i} | x_{iq} = 0; \boldsymbol{\pi}\right) = \prod_{k=1}^{q-1} \left(\frac{\pi_k}{\sum_{l=1}^{q-1} \pi_k}\right)^{I\left(\boldsymbol{x}_{(q-1)i} = \boldsymbol{e}_{(q-1)k}\right)}, \tag{5}$$

$\boldsymbol{e}_{(q-1)k}$ is a $(q-1)$-dimensional vector with a one at the $k$th coordinate and zeros at all other coordinates ($k \in [q-1]$). Note that $f\left(\boldsymbol{x}_{(q-1)i} | x_{iq} = 1; \boldsymbol{\pi}\right) = 0$ for all possible $\boldsymbol{x}_{(q-1)} \in \left\{\boldsymbol{e}_{(q-1)k}\right\}_{k=1}^{q-1}$. We now construct a marginal likelihood using expressions (1), (4), and (5):

$$\begin{aligned} L_f^{\boldsymbol{cd}}(\mathbf{X}_n) &= \sum_{i=1}^{n} \log f(\boldsymbol{x}_i; \boldsymbol{\pi}) + \sum_{i=1}^{n} I(x_{iq} = 0) \log f\left(\boldsymbol{x}_{(q-1)i} | x_{iq} = 0; \boldsymbol{\pi}\right) \\ &= \sum_{i=1}^{n} \sum_{k=1}^{q} I(\boldsymbol{x}_i = \boldsymbol{e}_k) \log(\pi_k) + \sum_{i=1}^{n} \sum_{k=1}^{q-1} I(\boldsymbol{x}_i = \boldsymbol{e}_k) \left[\log(\pi_k) - \log\left(\sum_{l=1}^{q-1} \pi_l\right)\right]. \end{aligned} \tag{6}$$

The PL function has corresponding pseudo-entropy expression

$$
\begin{aligned}
H(f) &= \sum_{\boldsymbol{x}\in\{\boldsymbol{e}_k\}_{k=1}^{q}} -\prod_{k=1}^{q}\pi_k^{I(\boldsymbol{x}_i=\boldsymbol{e}_k)}\left[\sum_{k=1}^{q} I(\boldsymbol{x}_i=\boldsymbol{e}_k)\log\pi_k\right] \\
&+ \sum_{\boldsymbol{x}_{(q-1)}\in\left\{\boldsymbol{e}_{(q-1)k}\right\}_{k=1}^{q-1}} -\prod_{k=1}^{q-1}\left(\frac{\pi_k}{\sum_{l=1}^{q-1}\pi_k}\right)^{I\left(\boldsymbol{x}_{(q-1)}=\boldsymbol{e}_{(q-1)k}\right)}\left[\sum_{k=1}^{q} I\left(\boldsymbol{x}_{(q-1)}=\boldsymbol{e}_{(q-1)k}\right)\log\left(\frac{\pi_k}{\sum_{l=1}^{q-1}\pi_k}\right)\right] \\
&+ \sum_{\boldsymbol{x}_{(q-1)}\in\left\{\boldsymbol{e}_{(q-1)k}\right\}_{k=1}^{q-1}} -0\times\log(0) \\
&= -\sum_{k=1}^{q}\pi_k\log\pi_k - \sum_{k=1}^{q-1}\left(\frac{\pi_k}{\sum_{l=1}^{q-1}\pi_k}\right)\log\left(\frac{\pi_k}{\sum_{l=1}^{q-1}\pi_k}\right). \qquad (7)
\end{aligned}
$$

Lines 2 and 3 of (7) are obtained by splitting the PMF (5) into the cases where $x_q = 1$ and when $x_q = 0$, respectively. Furthermore, we utilize the usual convention that $0 \times \log(0) = 0$ in order to obtain the final line of the equation (cf. MacKay, 2005, Sec. 2.4). Since $\pi_k > 0$ for all $k \in [q]$, we are guaranteed that each of the logarithms exist and are finite in $H(f)$ are finite. Thus, by Theorem (1), the maximization of (6) (with respect to the vector $\boldsymbol{\pi}$) yields a consistent estimator over the function class of form (4), in the sense of Section 2.

**Example 2.** *(Fully-visible Boltzmann Machine)* Let $\boldsymbol{X}_i \in \{0,1\}^q$ be a random variable ($i \in [n]$). We say that $\boldsymbol{X}_i$ arises from a fully-visible Boltzmann machine if the distribution of $\boldsymbol{X}_i$ can be characterized by the PMF

$$f(\boldsymbol{x}_i;\boldsymbol{\theta}) = \zeta^{-1}(\boldsymbol{\theta})\exp\left(\frac{1}{2}\boldsymbol{x}_i^\top\boldsymbol{M}\boldsymbol{x}_i + \boldsymbol{b}^\top\boldsymbol{x}_i\right), \qquad (8)$$

where $\boldsymbol{M} \in \mathbb{R}^{q\times q}$ is a symmetric matrix with $\operatorname{diag}(\boldsymbol{M}) = \boldsymbol{0}$ (the zero vector), $\boldsymbol{b} \in \mathbb{R}^q$, and $\boldsymbol{\theta}$ is a parameter vector that contains the unique elements of $\boldsymbol{M}$ and $\boldsymbol{b}$. Furthermore,

$$\zeta(\boldsymbol{\theta}) = \sum_{\boldsymbol{\xi}\in\{0,1\}^q}\exp\left(\frac{1}{2}\boldsymbol{\xi}^\top\boldsymbol{M}\boldsymbol{\xi} + \boldsymbol{b}^\top\boldsymbol{\xi}\right) \qquad (9)$$

is a normalization constant. Due to the exponentially increasing number of terms in (9), evaluating (8) becomes computationally intractable for any sufficiently large $q$. Thus, the maximization of the log-likelihood function that is obtained from (8) is infeasible in practice.

Using the notation from Example 1, we can derive the conditional PMFs

$$f\left(x_{ik}|\boldsymbol{X}_{(k)i}=\boldsymbol{x}_{(k)i};\boldsymbol{\theta}\right)=\frac{\exp\left(x_{ik}\boldsymbol{m}_k^\top\boldsymbol{x}_i+b_kx_{ik}\right)}{1+\exp\left(\boldsymbol{m}_k^\top\boldsymbol{x}_i+b_j\right)}, \tag{10}$$

where $\boldsymbol{m}_k$ is the $k$th column of $\boldsymbol{M}$ and $b_k$ is the $k$th element of $\boldsymbol{b}$, for each $k\in[q]$. Via the conditional PMFs (10), Hyvarinen (2006) considered the maximization of the log-PL function

$$\begin{aligned}L_f^{\boldsymbol{cd}}\left(\mathbf{X}_n;\boldsymbol{\theta}\right)&=\sum_{i=1}^{n}\sum_{k=1}^{q}\log f\left(x_{ik}|\boldsymbol{X}_{(k)i}=\boldsymbol{x}_{(k)i};\boldsymbol{\theta}\right)\\&=\sum_{i=1}^{n}\sum_{k=1}^{q}\left(x_{ik}\boldsymbol{m}_k^\top\boldsymbol{x}_i+b_kx_{ik}-\log\left[1+\exp\left(\boldsymbol{m}_k^\top\boldsymbol{x}_i+b_j\right)\right]\right),\end{aligned} \tag{11}$$

with respect to the parameter vector $\boldsymbol{\theta}$.

The PL function has corresponding pseudo-entropy expression

$$\begin{aligned}H\left(f\right)\;=\;&\sum_{k=1}^{q}\sum_{x_k\in\{0,1\}}\sum_{\boldsymbol{x}_{(k)\in\{0,1\}^{q-1}}}-\left[\frac{\exp\left(x_{ik}\boldsymbol{m}_k^\top\boldsymbol{x}_i+b_kx_{ik}\right)}{1+\exp\left(\boldsymbol{m}_k^\top\boldsymbol{x}_i+b_j\right)}\right.\\&\left.\times\left(x_{ik}\boldsymbol{m}_k^\top\boldsymbol{x}_i+b_kx_{ik}-\log\left[1+\exp\left(\boldsymbol{m}_k^\top\boldsymbol{x}_i+b_j\right)\right]\right)\right].\end{aligned}$$

Since the exponential function is always positive, the pseudo-entropy must exist and be finite for all valid parameter vectors $\boldsymbol{\theta}$. Thus, via application of Theorem 1, the maximization of the log-PL function (11) results in a consistent estimator for the parameter vector $\boldsymbol{\theta}$.

We note that consistency and asymptotic normality of the MPL estimator for the fully-visible Boltzmann machine were proved in Hyvarinen (2006) and Nguyen & Wood (2016b), respectively, when the log-PL function of form (11) is used. For the same estimation problem, Nguyen & Wood (2016a) proposed an efficient and convergent algorithm for the computation of the MPL estimator.

**Example 3.** *(Restricted Boltzmann Machine)* Let $\boldsymbol{X}_i\in\{0,1\}^q$ be a random variable that is observable($i\in[n]$). For each $i$, let $\boldsymbol{Y}_i\in\{0,1\}^r$ ($r\in\mathbb{N}$) be a latent variable, such that the joint PMF of $\boldsymbol{X}_i$ and $\boldsymbol{Y}_i$ is

$$f\left(\boldsymbol{x}_i,\boldsymbol{y}_i;\boldsymbol{\theta}\right)=\zeta^{-1}\left(\boldsymbol{\theta}\right)\exp\left(\boldsymbol{a}^\top\boldsymbol{x}_i+\boldsymbol{b}^\top\boldsymbol{y}_i+\boldsymbol{x}_i^\top\boldsymbol{M}\boldsymbol{y}_i\right) \tag{12}$$

where $\boldsymbol{M} \in \mathbb{R}^{q\times r}$, $\boldsymbol{a} \in \mathbb{R}^q$, $\boldsymbol{b} \in \mathbb{R}^r$, and $\boldsymbol{\theta}$ is a parameter vector that contains the scalar elements of $\boldsymbol{M}$, $\boldsymbol{a}$, and $\boldsymbol{b}$. Here,

$$\zeta\left(\boldsymbol{\theta}\right) = \sum_{\boldsymbol{\xi}\in\{0,1\}^q} \sum_{\boldsymbol{\upsilon}\in\{0,1\}^r} \exp\left(\boldsymbol{a}^\top\boldsymbol{\xi} + \boldsymbol{b}^\top\boldsymbol{\upsilon} + \boldsymbol{\xi}^\top\boldsymbol{M}\boldsymbol{\upsilon}\right),$$

is the form of the normalization constant.

Since $\boldsymbol{Y}_i$ is unobserved, for each $i$, we must marginalize (12) to obtain a PMF for $\boldsymbol{X}_i$. Via the law of total probability, the marginal PMF in terms of the observable random variable $\boldsymbol{X}_i$ is

$$f\left(\boldsymbol{x}_i;\boldsymbol{\theta}\right) = \omega^{-1}\left(\boldsymbol{\theta}\right)\exp\left(\boldsymbol{a}^\top\boldsymbol{x}_i + \sum_{k=1}^{r}\log\left[1+\exp\left(\boldsymbol{m}_k^\top\boldsymbol{x}_i + b_k\right)\right]\right), \tag{13}$$

where $\boldsymbol{m}_k$ is the $k$th column of $\boldsymbol{M}$ and $b_k$ is the $k$th element of $\boldsymbol{b}$. The normalization constant in (13) has the form

$$\omega\left(\boldsymbol{\theta}\right) = \sum_{\boldsymbol{\xi}\in\{0,1\}^q}\exp\left(\boldsymbol{a}^\top\boldsymbol{\xi} + \sum_{k=1}^{r}\log\left[1+\exp\left(\boldsymbol{m}_k^\top\boldsymbol{\xi} + b_k\right)\right]\right).$$

We say that $\boldsymbol{X}_i$ arises from a restricted Boltzmann machine if its distribution can be characterized in form (13).

Restricted Boltzmann machines have become an integral part of machine learning research, ever since their introduction in Smolensky (1986). This is due to their important roles as representational building blocks in deep belief networks as well as the availability of algorithms for the estimation of their parameters, including algorithms that are based on MPL estimation; see Bengio (2009, Sec. 5.3) and references therein for details.

Using the notation from Example 1 and writing $\text{softplus}\left(a\right) = \log\left(1+\exp a\right)$, for $a \in \mathbb{R}$, we can derive the conditional PMFs:

$$f\left(x_{ik}|\boldsymbol{X}_{(k)i} = \boldsymbol{x}_{(k)i};\boldsymbol{\theta}\right) = \frac{\exp\left[a_k x_{ik} + \sum_{l\neq k} a_l x_{il} + \sum_{j=1}^{r}\text{softplus}\left(m_{jk}x_{ik} + \sum_{l\neq j} m_{jl}x_{il} + b_j\right)\right]}{\sum_{\xi\in\{0,1\}}\exp\left[a_k\xi + \sum_{l\neq k} a_l x_{il} + \sum_{j=1}^{r}\text{softplus}\left(m_{jk}\xi + \sum_{l\neq j} m_{jl}x_{il} + b_j\right)\right]}, \tag{14}$$

where $a_l$ is the $l$th element of $\boldsymbol{a}$ and $m_{kl}$ is the $l$th element of $\boldsymbol{m}_k$, for each $k, l \in [q]$. Using (14), Marlin et al. (2010), Larochelle et al. (2012), and Yasuda et al. (2012) all considered the maximization of the log-PL function

$$L_f^{\boldsymbol{cd}}\left(\mathbf{X}_n;\boldsymbol{\theta}\right) = \sum_{i=1}^{n}\sum_{k=1}^{q}\log f\left(x_{ik}|\boldsymbol{X}_{(k)i} = \boldsymbol{x}_{(k)i};\boldsymbol{\theta}\right),$$

in order to obtain an estimator for $\boldsymbol{\theta}$. Since the exponential function and $\text{softmax}(\cdot)$ can both only return positive values, the corresponding pseudo-entropy

$$H(f) = \sum_{k=1}^{q} \sum_{x_k \in \{0,1\}} \sum_{\boldsymbol{x}_{(k)} \in \{0,1\}^{q-1}} -f\left(x_k | \boldsymbol{X}_{(k)} = \boldsymbol{x}_{(k)}; \boldsymbol{\theta}\right) \log f\left(x_k | \boldsymbol{X}_{(k)} = \boldsymbol{x}_{(k)}; \boldsymbol{\theta}\right)$$

must exist and be positive for all valid parameter vectors $\boldsymbol{\theta}$. Therefore, by Theorem 1, we obtain the consistency of the MPL estimator for the restricted Boltzmann machine. We note that the estimation of restricted Boltzmann machines can be viewed as parametric when $r$ is fixed, or semiparametric when $r$ is allowed to vary over some range of values.

## Acknowledgements

The author is grateful to the Associate Editor and two Reviewers for suggestions that have greatly improved the quality of the paper. The author is personally supported by Australian Research Council grant number DE170101134.